\documentclass[10pt, journal, a4paper, final, twoside, nofonttune]{IEEEtran}
\usepackage{amssymb,amsmath,amsthm,epsfig,graphics,subfigure, wrapfig, mhchem, hyperref}
\usepackage[english]{babel}

\begin{document}

\title{Device for measuring the plant physiology\\ and electrophysiology}
\vspace{3mm}
\author{Serge Kernbach\\[2mm]
{\small CYBRES GmbH, Research Center of Advanced Robotics and Environmental Science,\\Melunerstr. 40, 70569 Stuttgart, Germany, \textit{serge.kernbach@cybertronica.de.com}}
}

\date{}
\maketitle
\thispagestyle{empty}

\begin{abstract}
This paper briefly describes the device -- the phytosensor -- for measuring physiological and electrophysiological parameters of plants. This system is developed as a bio-physiological  sensor in precise agriculture, as a tool in plant research and environmental biology, and for plant enthusiasts in smart home or entertainment applications. The phytosentor measures main physiological parameters such as the leaf transpiration rate, sap flow, tissue conductivity and frequency response, biopotentials (action potentials and variation potentials), and can conduct electrochemical impedance spectroscopy with organic tissues. Soil moisture and temperature, air quality (\ce{CO_2}, \ce{NO_2}, \ce{O_3} and other sensors on \ce{I^2C} bus), and general environmental parameters (light, temperature, humidity, air pressure, electromagnetic and magnetic fields) are also recorded in real time. In addition to phytosensing, the device can also perform phytoactuation, i.e. execute electrical or light stimulation of plants, control irrigation and lighting modes, conduct fully autonomous experiments with complex feedback-based and adaptive scenarios in robotic or biohybrid systems. This article represents the revised and extended version of original paper and includes some descriptions and images from the FloraRobotica and BioHybrids projects.
\end{abstract}

\section{Introduction}

\begin{figure}[htp]
\centering
{\includegraphics[width=0.4\textwidth]{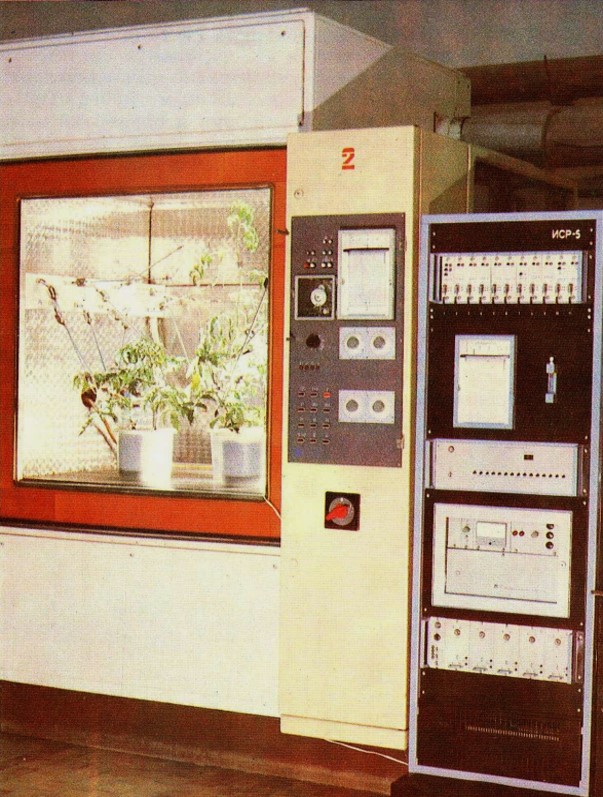}}
\caption{\small BIOTRON, the system developed in the Institute of Genetics in Academy of Sciences of Moldova in 80s and 90s. This system was used in agricultural purposes -- optimization of production, fast tests of genetic hybrids. \label{fig:biotron}}
\end{figure}

The system for measuring physiological and electrophysiological reactions of phyto-objects is intended for analy\-zing and monitoring responses of phyto-objects to external stimuli. Original development was inspired by works of S.N.Maslobrod \cite{maslobrod04S} and the developed BIOTRON\footnote{Developed in the Institute of Genetics in Academy of Sciences of Moldova in 80s and 90s.} system, see Fig.\ref{fig:biotron}. The BIONRON was used for plant research in agriculture, e.g. for optimization of production or fast tests of genetic hybrids, and involved different sensors and computational units. Next developmental steps have been done within the FloraRobotica \cite{FloraRobotica} and BioHybrids projects \cite{Biohybrids} with focus on different sensors and physiological processes in plants \cite{hamann2017flora}, \cite{Kernbach18Yeasten}.

The phytosensing system is designed for professional and amateur purposes, as sensors in precise agriculture, as a tool for biological laboratories, and for plant enthusiasts. In the professional use this system can be employed to study the electrical reaction of plants, as phyto- or bio-sensors, in the bio-hybrid and robotic systems, for systems of mixed reality, smart home applications, for biotechnological processes based on phyto-objects. Additional algorithms for data analysis and actuation can be implemented with the provided script language. In amateur applications the system has a variety of potential usages, for example, in school or university courses on biology, for studying ‘the electrical language of plants’, or to carry out experiments in the context of Cleve Backster \cite{Backster68} and Peter Tompkins works with the ‘Secret life of plants’ \cite{Tompkins73}.

In addition to phytosensing, the device can also perform phytoactuation, i.e. execute real-time data processing and decision making for operating different actuators. This functionality is useful for electrical or light stimulation of plants (blue/red light stimulation), optimizing irrigation and lighting modes, conducting fully autonomous experiments with complex feedback-based and adaptive scenarios in biohybrid systems. It possesses built-in semiconductor relays (switches) e.g. for RGB LEDs or water pumps, and can actuate external USB relays. The system implements dynamical mapping between embedded signal detectors, numerical processors and actuators with reactive, probabilistic and homeostatic decision mechanisms. The phytosensor is a part of \hbox{CYBRES} biohybrid interface devices with various sensors that measure the plant physiology and environmental parameters\footnote{see \url{www.youtube.com/watch?v=_xfKOYOpNU4}.}.

In the extended configuration with external mini-PC, the system can be configured as an indicator of biological pathogenicity of the environment. It can measure separate parameters (e.g. \ce{CO_2}, \ce{NO_2}, \ce{O_3} or electromagnetic emission level) or a ‘complex biological pathogenicity’ – a combination of different environmental parameters that affect the metabolism. The system needs about one-two weeks to adapt to the plant, its signals and to calculate correlations with other sensors. For instance, in this mode the green LED indicates a normal signal level of the plant, yellow -- the deviation of current parameters from a ‘normal’ state, red -- the critical state. The main purpose of this operational mode is to attract attention to possible biological pathogenicity on early stage.

In all configurations, the device provides access to measured data and allows their algorithmic, mathematical and statistical analysis. This functionality is useful for optimization of plant’s productive functionality, for using plants as biosensors, including weak signals, integration into systems of mixed reality, smart home applications as well as for scientific research of the world of plants. In particular, it can be used to explore the learning and adaptation processes in vegetable and biohybrid systems based on plants \cite{Gagliano16}.

Functions with two channels enable performing high-resolution differential measurement. Real-time data (with time stamps) can be recorded into the internal flash memory, transmitted via USB or (with optional mini-PC) are accessible via WiFi/Ethernet/Bluetooth on any mobile device/PC. Data can be written as html-pages for the online data plot in internet. The needle- or surface- electrodes are used to receive signals from plants. The system is fully autonomous and can operate without user intervention.

The system measures the following parameters (depending on used electrodes in basic and pro versions):

\begin{enumerate}
	\item 
    \textbf{Biopotentials} (2 channels, Ag99 needles), used for:\\
       - measuring fast electrical response to different \mbox{stimuli};\\
       - tracking periodical activities of the plant;\\
       - interactions with plants based on e.g. mechanical or thermal stimuli;\\
       - other bio-electric approaches;
	
	\item
    \textbf{Tissue impedance} sensors (2 channels, Ag99 needles), used for:\\
       - measuring conductivity of organic tissues at different frequencies;\\
       - frequency shift in tissue response;\\
       - electrochemical sap flow measurements;\\
       - electrical stimulation;\\
			 - electrochemical impedance spectroscopy of organic tissues and fluids;\\
      - other ionic interfaces to organic/living materials/objects;
			
	\item	\textbf{Leaf transpiration} (single channel);
	
	\item	\textbf{Sap flow} (single channel) based on electrochemical or thermal principles;
					
	\item \textbf{Soil moisture and temperature};				
	
	\item  \textbf{Environmental parameters}: 3D accelerometer/magnetometer, EM power meter (450kHz-2.5GHz), air temperature and humidity, air pressure, light;
	
  \item  Supported \textbf{external sensors}: \ce{CO_2}, \ce{O_3}, \ce{No_2}, ion-selective electrodes, any sensors with \ce{I^2C} interface.				
					
\end{enumerate}

\begin{figure*}[htp]
\centering
{\includegraphics[width=1\textwidth]{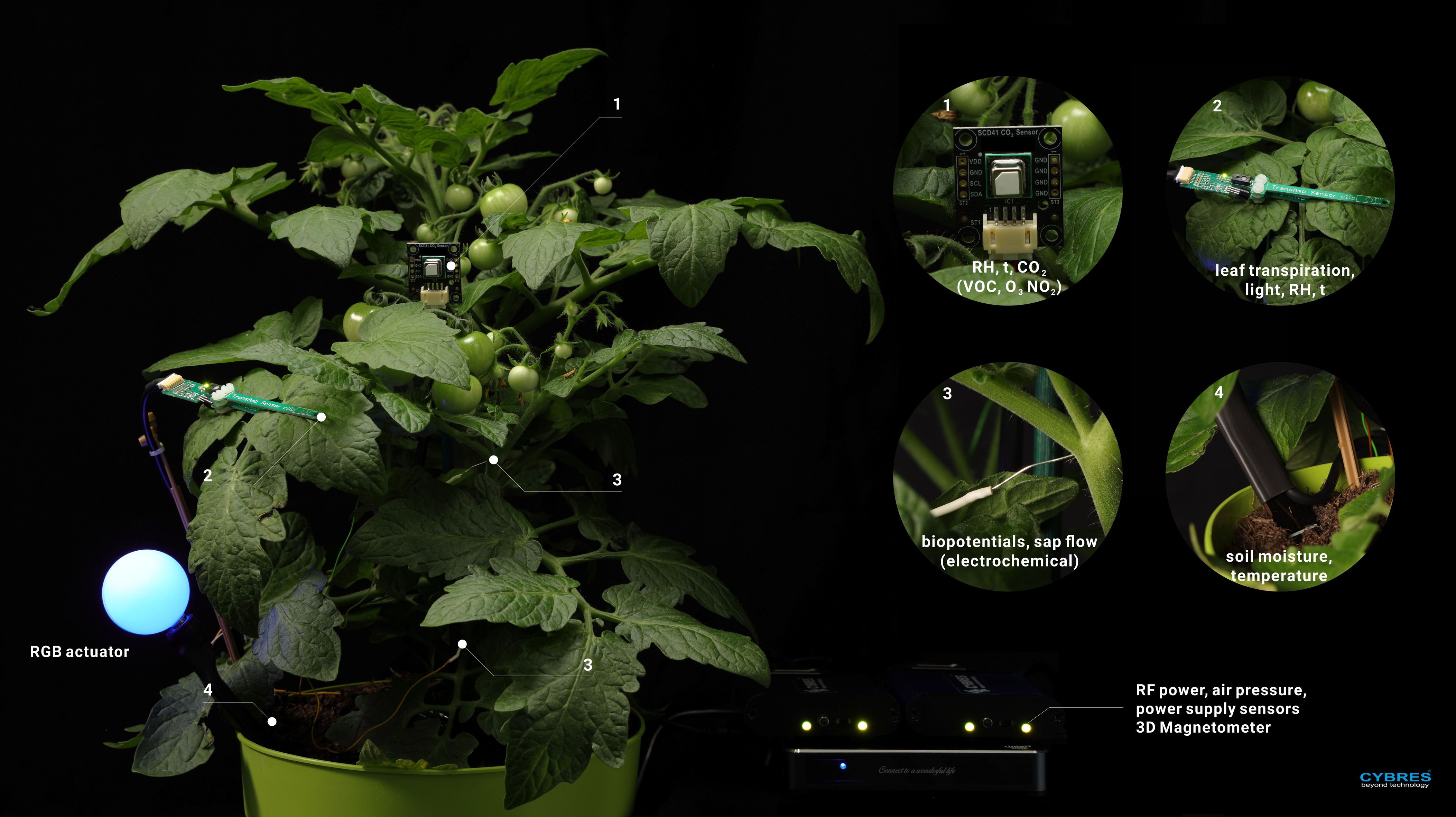}}
\caption{\small Image of the phytosensor with tomato plant, see description in text. \label{fig:structure1}}
\end{figure*}

Several images of the system are shown in Figs.\ref{fig:structure1}, \ref{fig:PhytosensorSensors}, \ref{fig:structure3}, \ref{fig:Phytosensor}. Main electrical parameters are summarized in Table \ref{tab:electricParameters} 

\begin{table}[htp]
\centering
\caption{\small Brief overview of main electrical parameters of the phytosensing system. \label{tab:electricParameters}}
\fontsize {9} {10} \selectfont
\begin{tabular}{
p{6.0cm}@{\extracolsep{3mm}}
p{3.0cm}@{\extracolsep{3mm}}
}
Resolution of biopotentials (input current) & $\pm$64nV ($\pm$70pA) \\
The potential for electrical stimulation and measurement of tissue conductivity & $\pm$10mV -- $\pm$1B \\
The frequencies of electrical stimulation and measurement of tissue conductivity & 8Hz -- 0.3MHz ~~~(max. 0.65MHz)\\
Intervals between measurements & 0.1sec -- 100 seconds\\
Internal flash memory & 512MB \\
Analysis of the frequency shift & FFT, FRA \\
Power supply & USB, 5V \\
\end{tabular}
\end{table}

\section{Electrodes and sensors}

\begin{figure*}[htp]
\centering
{\includegraphics[width=1\textwidth]{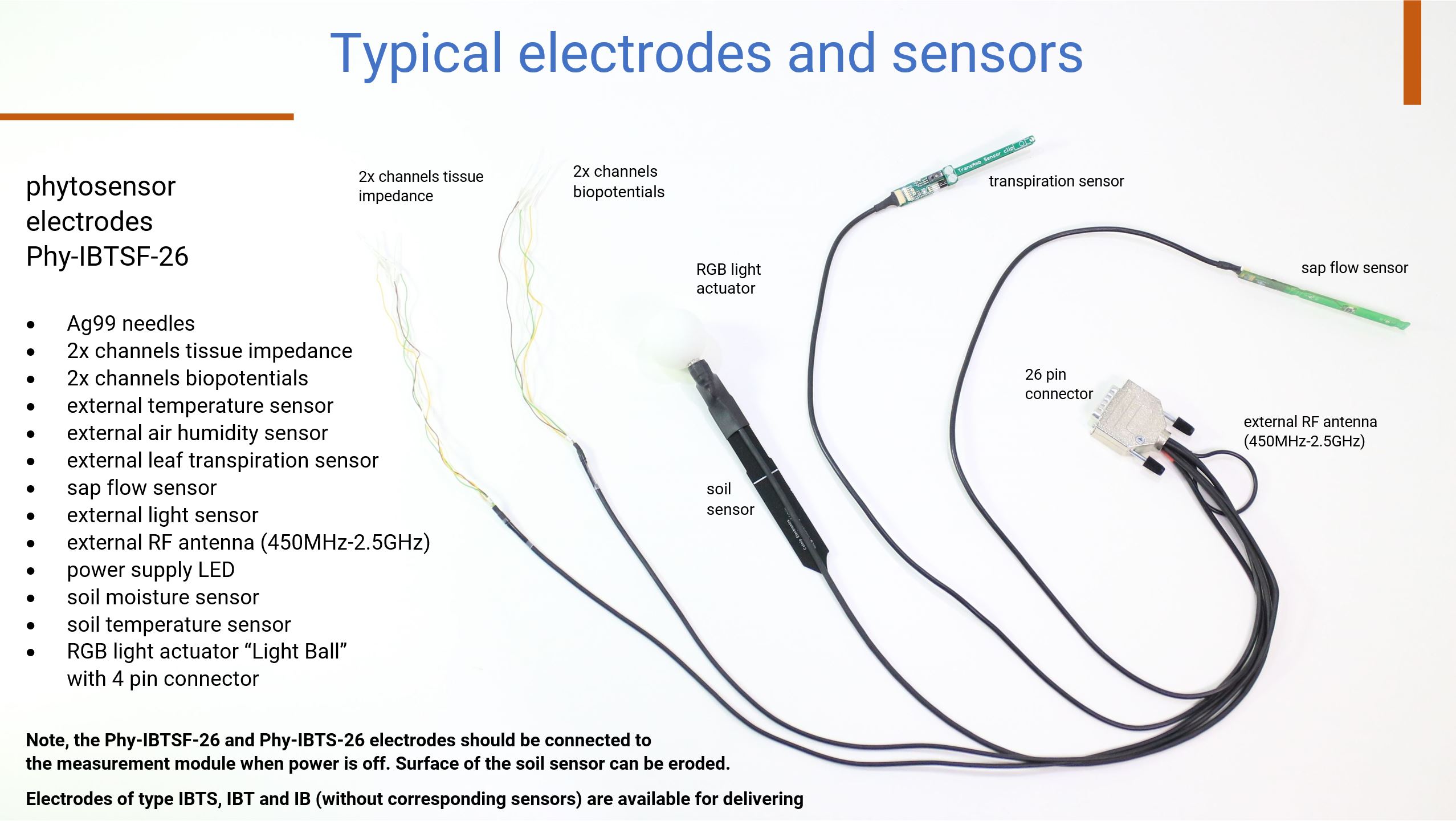}}
\caption{\small Example of electrodes in the phytosensing system. \label{fig:electrodes}}
\end{figure*}

The phytosensing system is provided with different electrodes, as shown in Fig. \ref{fig:electrodes}:

\begin{itemize}
 \item   Bio-potential/tissue-conductivity electrodes (Ag99);
 \item   The leaf transpiration sensor;
 \item   The senor panel with several environmental sensors (air temperature, air humidity, light);
 \item   The soil moisture/soil temperature sensor;
 \item   The light-ball actuator;
 \item   Stainless steel electrodes for trees/shrubs;
 \item   Electrochemical/thermal sap flow sensor.
\end{itemize}

Additionally, the system has optional components:

\begin{itemize}
 \item  IP65/66 water protected package for outdoor applications;
 \item  The MU3.4 module with integrated sensors (3D accelerometer/magnetometer, EM power meter (450kHz-2.5GHz), air pressure) and USB 2.0 connectivity;
 \item  The mini-PC for advanced connectivity with WiFi/Ethernet/Bluetooth;
 \item USB server for WiFi/Etherner connectivity without PC;
 \item USB relay for switching 110/220V load (motors, lamp, pumps, etc.).
\end{itemize}

\section{(Electro-)physiological measurements}

The phytosensing system has two high-impedance inputs for measuring potentials (voltage) in terms of action potentials \cite{10.3389/fpls.2019.00082} or variation potentials \cite{Davies06} that can be used for: 1) bio-potential measurements in the phytosensor configuration; 2) general purpose two-channel high-resolution voltage measurements and logging. Usage in 1) or 2) requires different order and timing of sampling. Default configuration is 1), where first the voltage channels are sampled, then impedance channels are sampled, and finally all other sensors are sampled. This order of sampling provides a low level of distortions for sensing bio-potentials.

\begin{figure}[htp]
\centering
\subfigure{\includegraphics[width=0.51\textwidth]{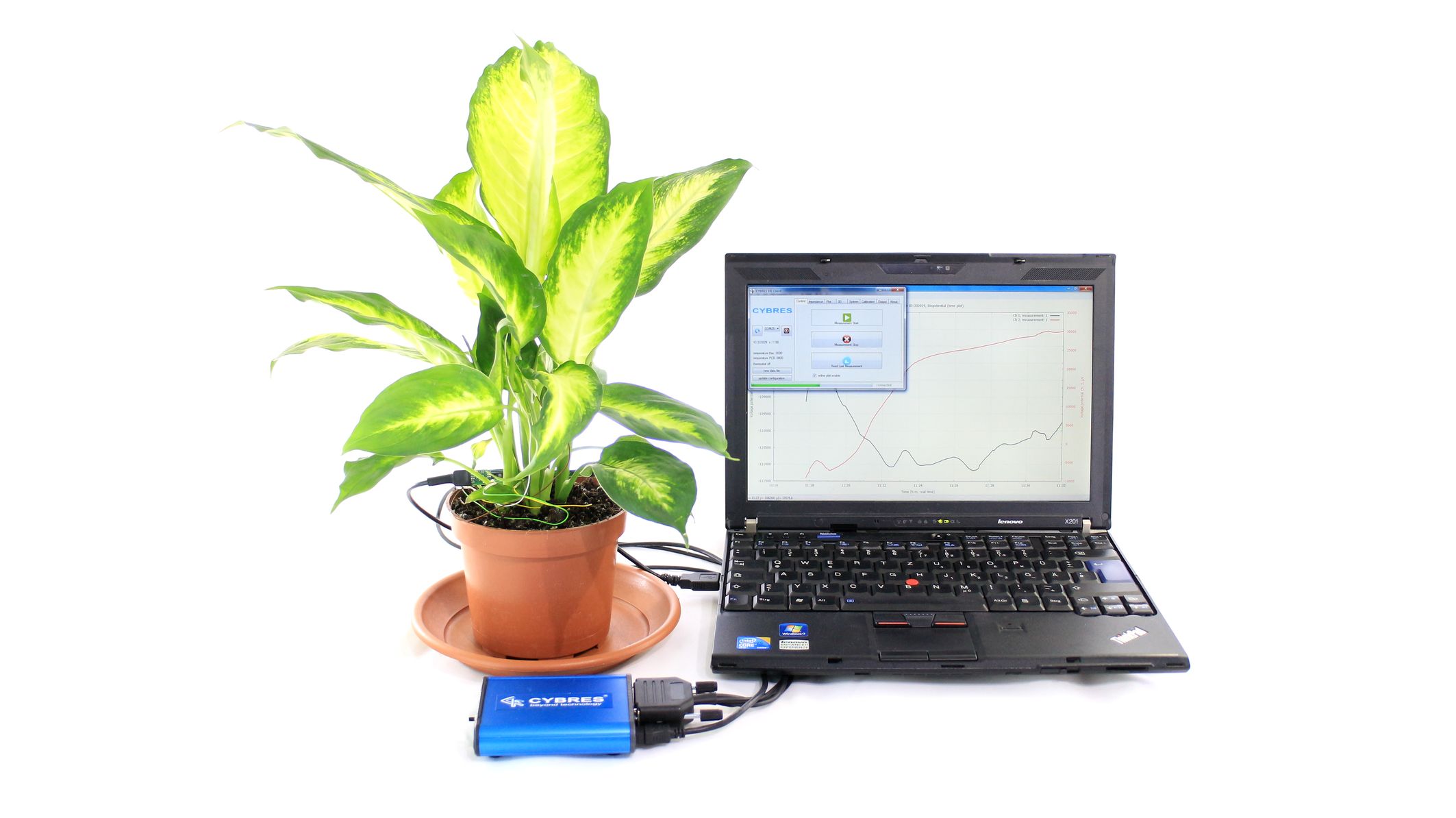}}
\caption{\small The phytosensor connected to a laptop. \label{fig:PhytosensorSensors}}
\end{figure}

\begin{figure*}[htp]
\centering
{\includegraphics[width=1\textwidth]{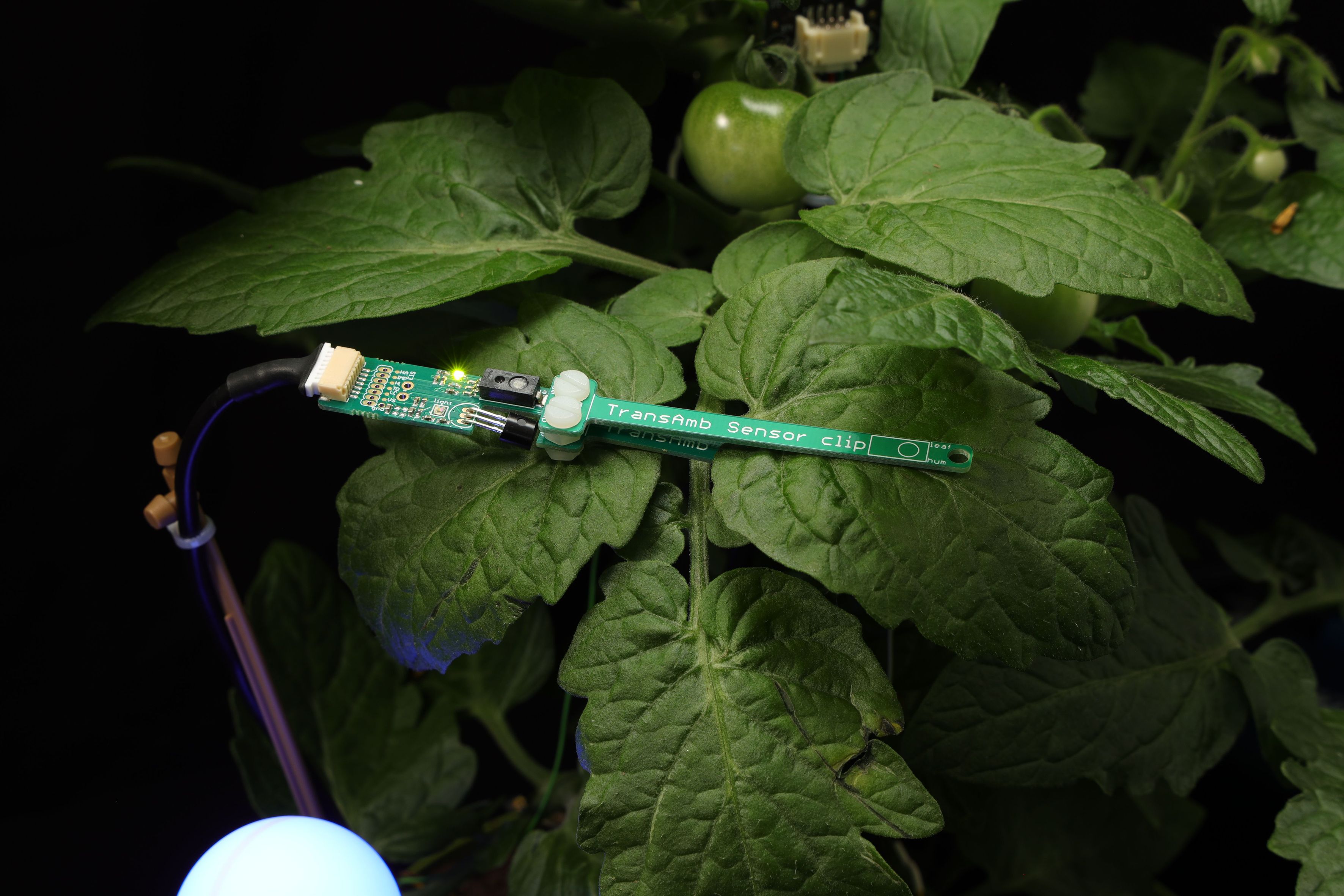}}
\caption{\small Phytosensor: leaf transpiration sensor, see description in text. \label{fig:structure3}}
\end{figure*}

\begin{figure*}[htp]
\centering
{\includegraphics[width=1\textwidth]{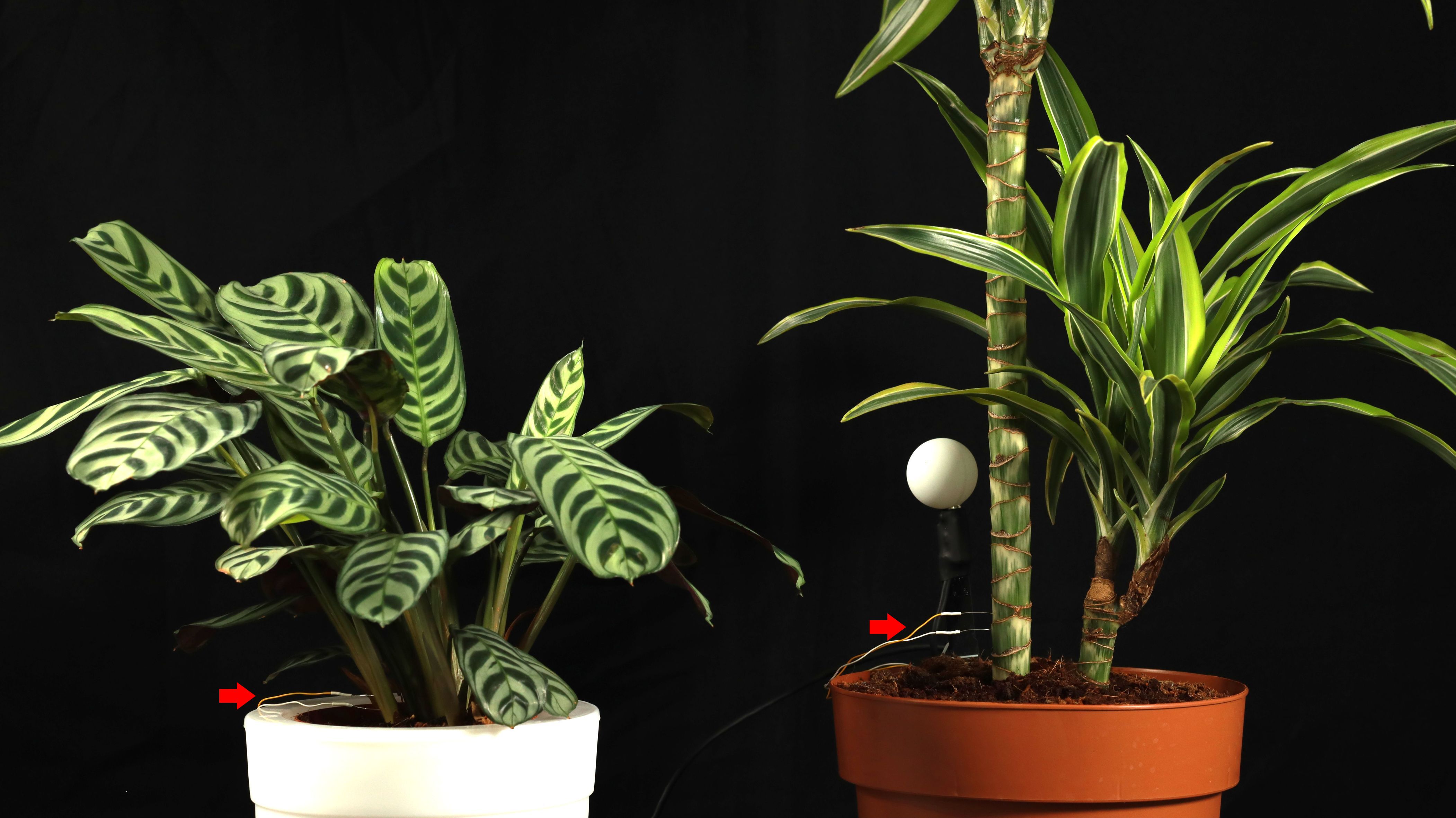}}
\caption{\small Phytosensor: electrochemical sap flow sensor, see description in text. \label{fig:structure4}}
\end{figure*}

The potential input has a high input impedance (input bias current is about $\pm$70pA), this enables sensing of bio-potentials in electrophysiological applications \cite{Chatterjee15}, e.g. plants, organic tissues, microbial fuel cell (MFC), microorganisms and similar applications. The current flowing through the test system, can be used not only for tissue impedance measurements, but also for electro-stimulation purposes. Essential advantage of this scheme is the flexible/variable frequency of stimulation, and automatic timing (e.g. a stimulation pulse every 10 sec.). Physiological measurements with plants include data from specific sensors such as the leaf transpiration, see Fig. \ref{fig:structure3}, or sap flow, see Fig. \ref{fig:structure4}, sensors. Several applications require external PC/laptop for data analysis and user interaction purposes, see Fig. \ref{fig:PhytosensorSensors}.

\section{Tissue impedance measurements}

Tissue impedance measurement is a well-known laboratory technique in biological research \cite{Cohn1970}, e.g., in the analysis of DNA or structure of tissues, the analysis of surface properties and control of materials. In the phytosensing system it is used for four applications: electrochemical sap flow sensor, detection of frequency shift in organic tissues, electrochemical impedance spectroscopy (EIS), and general purpose tissue impedance measurement \cite{10.3389/fbioe.2019.00474}. This method consists in applying a small AC voltage into a test system and registering a flowing current. Based on the voltage and current ratios, the electrical impedance $Z(f)$ for a harmonic signal of frequency $f$ is calculated. Measured data are fitted to the model of considered system and allow identifying a number of physical and chemical parameters.

A common approach consists in analyzing the frequency response (frequency response analysis -- FRA) of the $V_I$ signal, which is based on the discrete Fourier transform (DFT) and synthesis of ideal frequencies. This method is sometimes called as the single point DFT. The digitized time signal $V_I(k)$ with $N$ samples is converted to a frequency signal, containing real $Re^{FRA}(V_I)$ and imaginary $Im^{FRA}(V_I)$ parts
\begin{eqnarray}
\label{eq1}
Re^{FRA}(V_I(f))+iIm^{FRA}(V_I(f))=\nonumber\\
=\frac{1}{N}\sum_{k=0}^{N-1} V_I(k) \left[\cos(\frac{2\pi fk}{N})-i \sin(\frac{2\pi fk}{N})\right].
\end{eqnarray}
The required by FRA period-stable detection of $V_I(k)$ signal is implemented in hardware in the system-on-chip. The FRA magnitude $M(f)$ and phase $P(f)$ of the signal are calculated as
\begin{eqnarray}
\label{eq4}
M(f)&=&\sqrt{Re^{FRA}(V_I(f))^2+Im^{FRA}(V_I(f))^2},\\ P(f)&=&\tan^{-1}(Im^{FRA}(V_I(f))/Re^{FRA}(V_I(f))).
\end{eqnarray}

Additionally, the differential EIS meter uses a phase-amplitude detection of excitation and response signals. The RMS values $V_V^{RMS}$ and $V_I^{RMS}$ (it needs to remember that these signals are frequency $f$ dependent, i.e. $V_V^{RMS}(f)$ and $V_I^{RMS}(f)$) are calculated as

\begin{eqnarray}
\label{eq5}
V_I^{RMS}(f)&=& \sqrt {\sum_{k=0}^{N-1} \frac{1}{N} \left(V_I^{f}(k)\right)^2},\\
V_V^{RMS}(f)&=& \sqrt {\sum_{k=0}^{N-1} \frac{1}{N} \left(V_V^{f}(k)\right)^2}.\\
\end{eqnarray}
They are used for calculating so-called 'RMS resistivity'
\begin{eqnarray}
\label{eq5ab}
M^{RMS}(f)&=&\frac{V_V^{RMS}(f)}{V_I^{RMS}(f)}.
\end{eqnarray}
$M^{RMS}(f)$, calculated from RMS values, corresponds to the magnitude of impedance $M(f)$, calculated by FRA. The $V_I^f(k)$, $V_V^f(k)$ samples allow calculating two other values --
the correlation $C(f)$ and phase $P^C(f)$ (based on the lock-in phase detector for harmonic signals)
\begin{eqnarray}
\label{eq5a}
C(f)&=&\frac{1}{N}\sum_{k=0}^{N-1} V_I^f(k) V_V^f(k),\\
P^C(f)&=&\frac{180}{\pi}\cos^{-1}(\gamma(f) C(f)),
\end{eqnarray}
where $\gamma(f)$ is a $f$-dependent amplitude-based coefficient, detected in $V_I$, $V_V$ signals. The $P^C(f)$ is equivalent to $P(f)$, calculated by FRA. Considering \mbox{$M^{RMS}(f)$ and $P^C(f)$}, the value of $C(f)$ contains both the phase and amplitude characteristic of $V_I$, $V_V$ signals and thus it is the most appropriate as a single output value. The analysis allows identifying five main parameters:
\begin{enumerate}
  \item the differential amplitude of excitation and response signals (these values are included in all amplitude characteristics obtained by FRA, RMS and correlation approaches);
  \item the differential magnitude/conductivity;
  \item the differential phase;
  \item the differential correlation of excitation and response signals;
  \item variations of electrochemical stability of samples in time, frequency and time-frequency domains.
\end{enumerate}
The signal scope mode is also suitable for distortion analysis, e.g. the frequency shift in organic tissues. 

\section{Environmental measurements}

Important feature of the phytosensing system is its capability to measure different environmental parameters during 'main measurements'. There are fixed sensors installed on the PCB, additional sensors installed in electrodes, and replaceable sensors with digital (\ce{I^2C}) and analog interface in 26 pin connectors. All these sensors can be turned on/off. Data channels from these sensors can be used for numerical and statistic calculations in real time.

The device is equipped with several temperature sensors (temperature of the thermostat and electronic module, the control of power supply (used to monitor interferences on the power line), control of the thermostat (used to control the energy level supplied to the measuring part of the device), 3D magnetometer (used to control the static magnetic field during experiments) and 3D accelerometer (used to control mechanical impacts). The system also have the 450Mhz-2.5Ghz RF power meter. 

Measuring module supports an external high resolution temperature sensor (Texas Instruments LM35CA), connected to the side connector. It has a typical absolute accuracy of $\pm0.2^\circ$C, typical nonlinearity of $\pm0.15^\circ$C and the conversion factor V/$^\circ$C of \mbox{+10mV/$^\circ$C} (see more the Datasheet 'LM35 Precision Centigrade Temperature Sensors'). With the ADC resolution of 22 bit, this sensor provides a resolution of relative temperature measurements $<0.001^\circ$C. The sensor is useful for monitoring the temperature of objects around the system or the air temperature.

\section{Reconfigurable hardware implementation}

The implemented hardware is shown in Fig. \ref{fig:PhytosensorSensors} and Fig. \ref{fig:Phytosensor}. It consists of the main measurement unit (MU), electrodes and a statistical server (mini PC or laptop). The  system also contains a number of sensors for measuring environmental conditions. Example of replaceable electrodes (different optical excitation, volume, combination with magnetic excitation and others) with are shown in Fig. \ref{fig:PhytosensorSensors}.

Important feature of the phytosensing system is a combination with the electrochemical impedance spectrometer that enables a cost-effective solution for fluidic and biological analysis by means of hardware reconfiguration. Different reconfigurable hardware and software options are enlisted in Table \ref{tab:deviceVersions}.

\begin{table*}[htp]
\centering
\caption{\small Reconfigurable hardware and software options available for the phytosensing system. \label{tab:deviceVersions}}
\fontsize {9} {10} \selectfont
\begin{tabular}{
p{0.3cm}@{\extracolsep{3mm}}
p{6.0cm}@{\extracolsep{3mm}}
p{8.0cm}@{\extracolsep{3mm}}
p{1.5cm}@{\extracolsep{3mm}}
}\hline
N	& Device versions &	Hardware &	Software \\ \hline
1	& (\textbf{reconfigurable}) phytosensor basic	& 1 channel biopotentials \& impedance &	 enabled \\
2	& (\textbf{reconfigurable}) phytosensor advanced &	+ TransAmb sensor (leaf transpiration) + 2 channels biopotentials \& impedance &	enabled \\
3	& (\textbf{reconfigurable}) phytosensor full &	+ sap flow sensor	& enabled \\
4	& (\textbf{reconfigurable}) EIS &	+ 2x EIS open electrodes, + fluidic/environ. t sensors, + excitation spectroscopy & activation code \\
5	& (\textbf{embedded}) EIS &	different device with thermostat, + RGB/IR excitation spectroscopy & enabled \\
6	& (\textbf{embedded} \& \textbf{reconfigurable}) biosensor	& + fermentation module, + RGB/IR excitation spectroscopy	& enabled \\
\hline
\end{tabular}
\end{table*}

The measuring unit (MU) -- MU20, MU31, MU32, MU33, MU34, MU34T and MU40 -- is based on the 32-bit ARM processor with a real-time operating system (the MU RTOS). It has accurate analog-to-digital and digital-to-analog converters, an internal non-volatile memory, real time clock, blocks of low-pass filters and a number of additional sensors. The measuring unit is characterized by a low noise level. The MU platform enables connection of different real-world actuators and is used in different (e.g. European) research and industrial projects.

\begin{figure*}[htp]
\centering
\subfigure{\includegraphics[width=1.\textwidth]{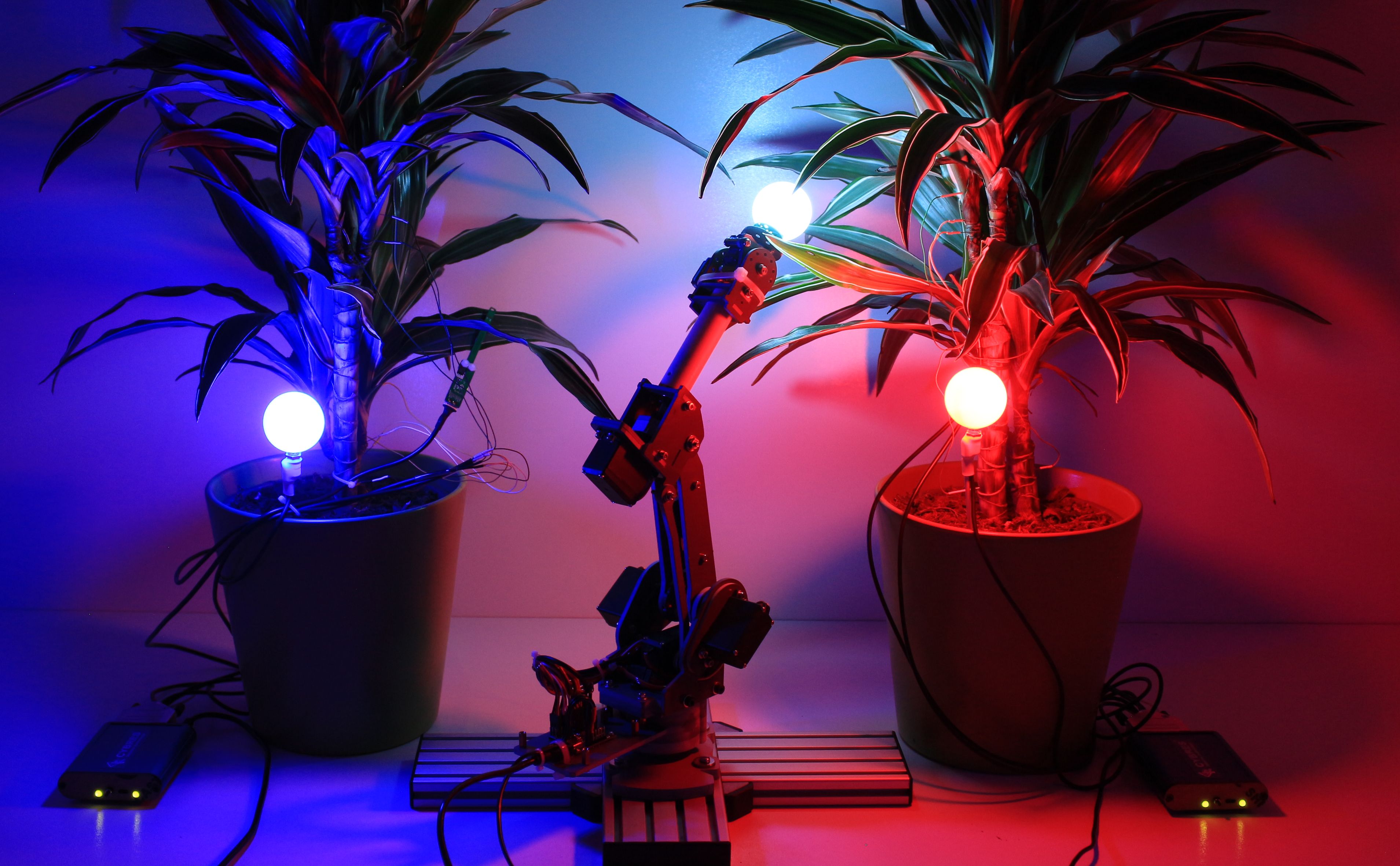}}
\caption{\small Phytoactuation: plants actuate robot manipulator and two light sources, see the following video
\url{www.youtube.com/watch?v=AqRtfHuPLH0}. 
\label{fig:detectors}}
\end{figure*}

\section{Software implementation}

\begin{figure}[htp]
	\centering
	\includegraphics[width=0.45\textwidth]{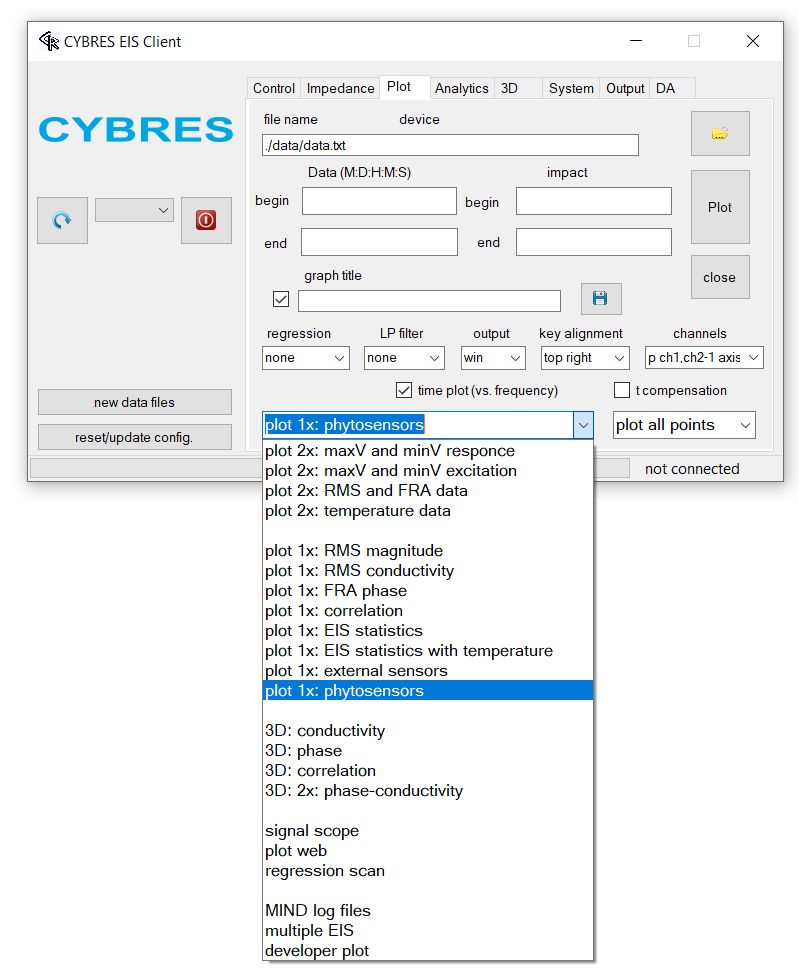}
	\includegraphics[width=0.45\textwidth]{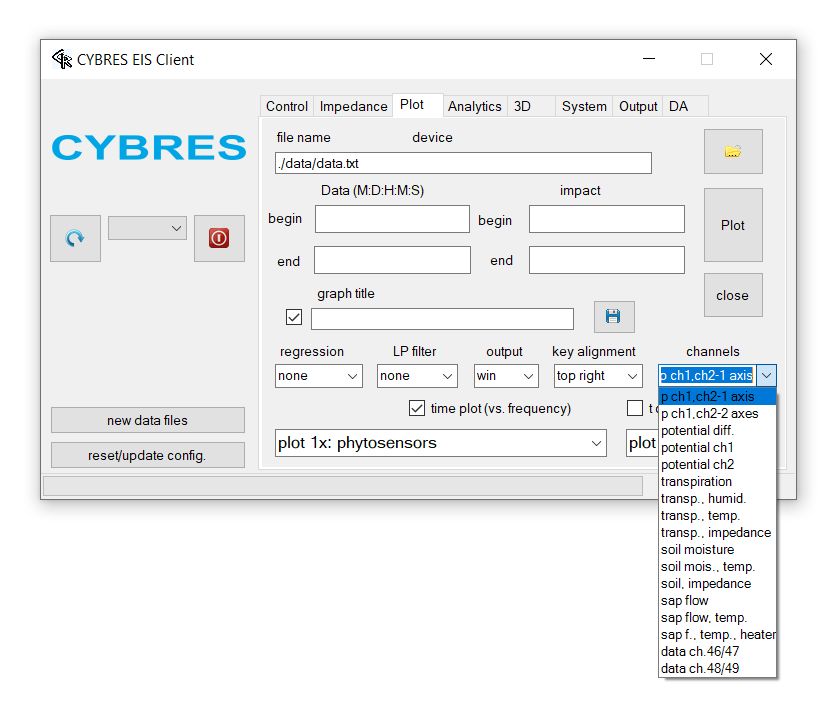}
	\includegraphics[width=0.45\textwidth]{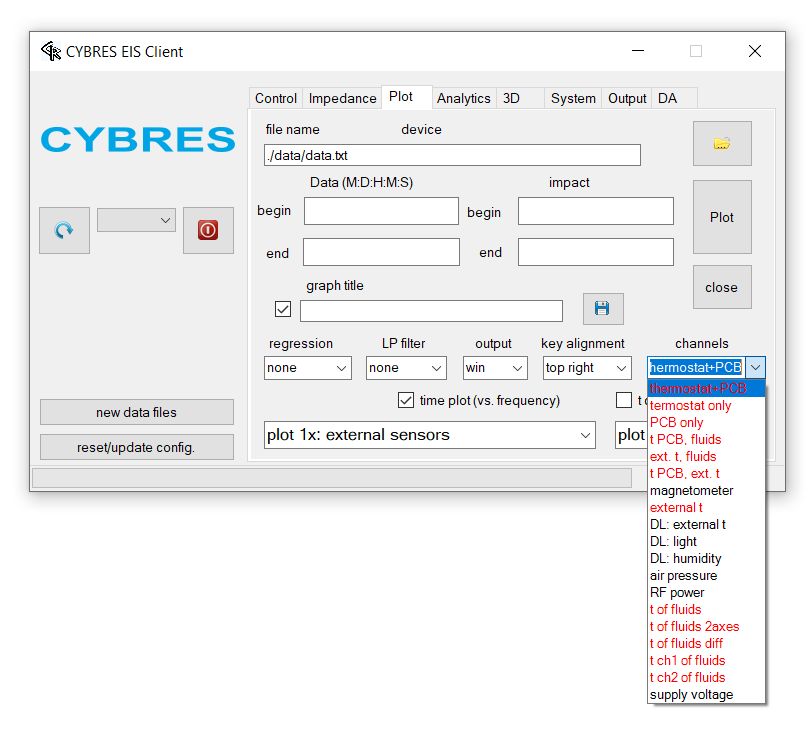}
    \caption{Examples of client program interface for phytosensing applications: main menu, selection of measurement modes and external sensors.\label{fig:cleint1}}
\end{figure}

Software includes the real-time operating system -- MU OS -- developed by CYBRES for prog\-rammable-systems-on-chip (PSOC) and the client-level software. The MU client performs all main tasks related to device and file management, handling time and excitation. The client program operates with gnuplot and DA (detector-actuator) scripts. Gnuplot scripts represents the third software level and is responsible for graphical utilities, 3D/4D plot and regression functionality. Finally, DA scripts handle statistical data processing, sensor-fusion functionality, perform actuator control and multi-device management. Gnuplot and DA scripts are open for users and can be customized for particular purposes. Examples of client interface, related to the phytosensing applications, are shown in Fig. \ref{fig:cleint1}.

\section{Example of data plots and test measurements}

\begin{figure}[ht]
	\centering
	\includegraphics[width=0.49\textwidth]{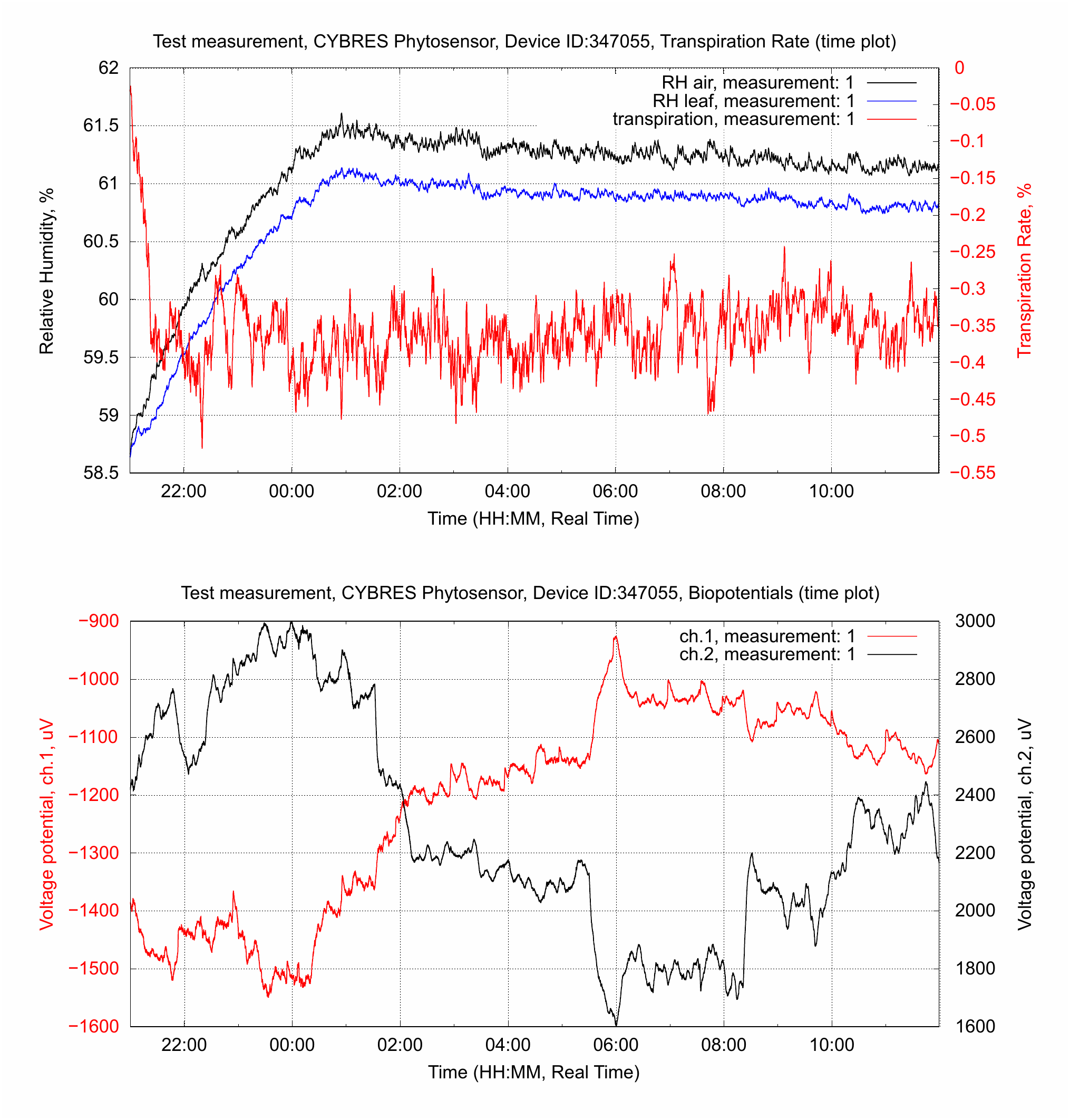}
    \caption{Test measurements: transpiration and bio-potentials.\label{fig:testMeasurement1}}
\end{figure}

\begin{figure}[htp]
	\centering
	\includegraphics[width=0.49\textwidth]{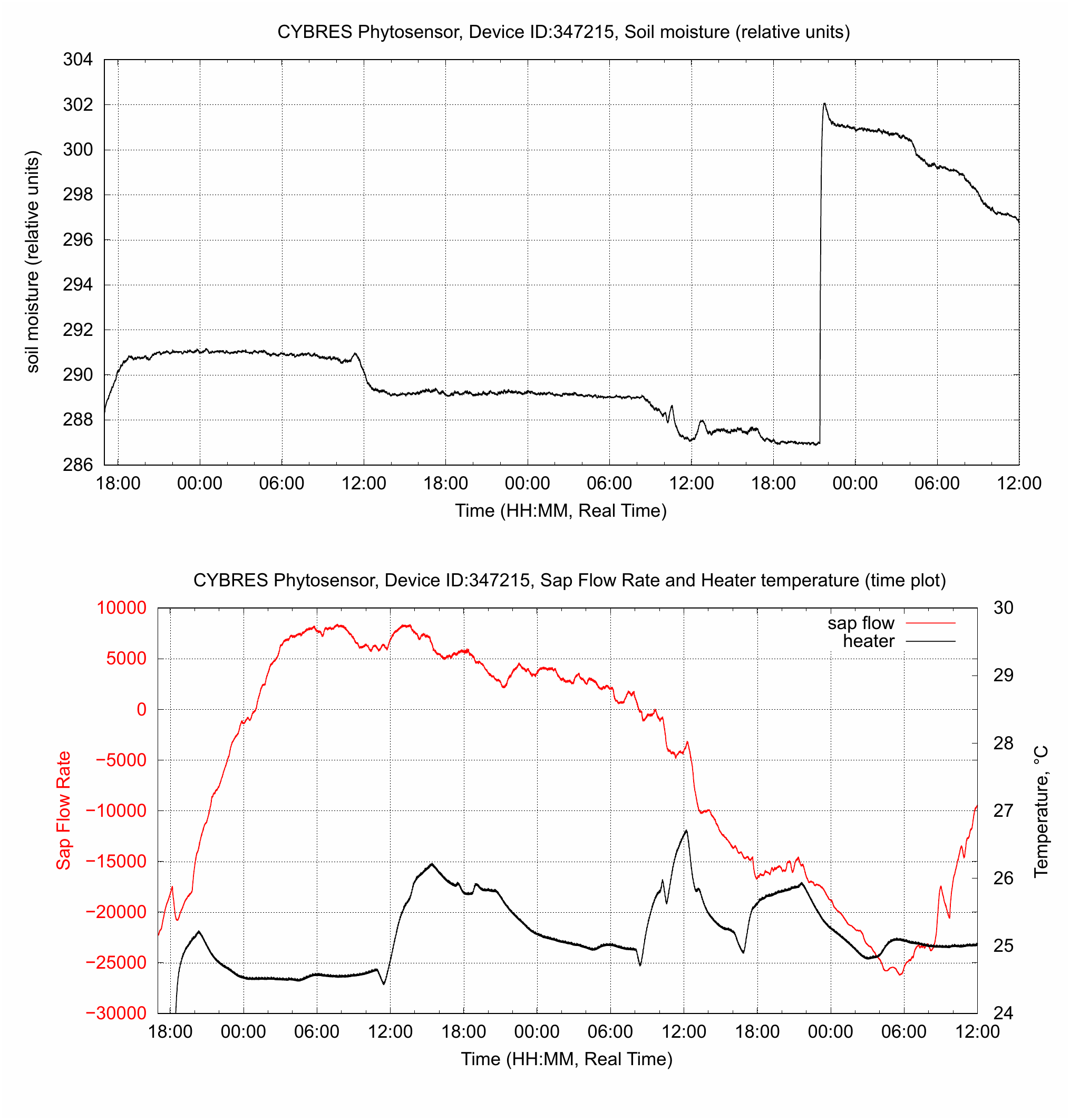}
    \caption{Test measurements: soil moisture and sap flow sensors.\label{fig:testMeasurement2}}
\end{figure}

\begin{figure}[htp]
	\centering
	\includegraphics[width=0.49\textwidth]{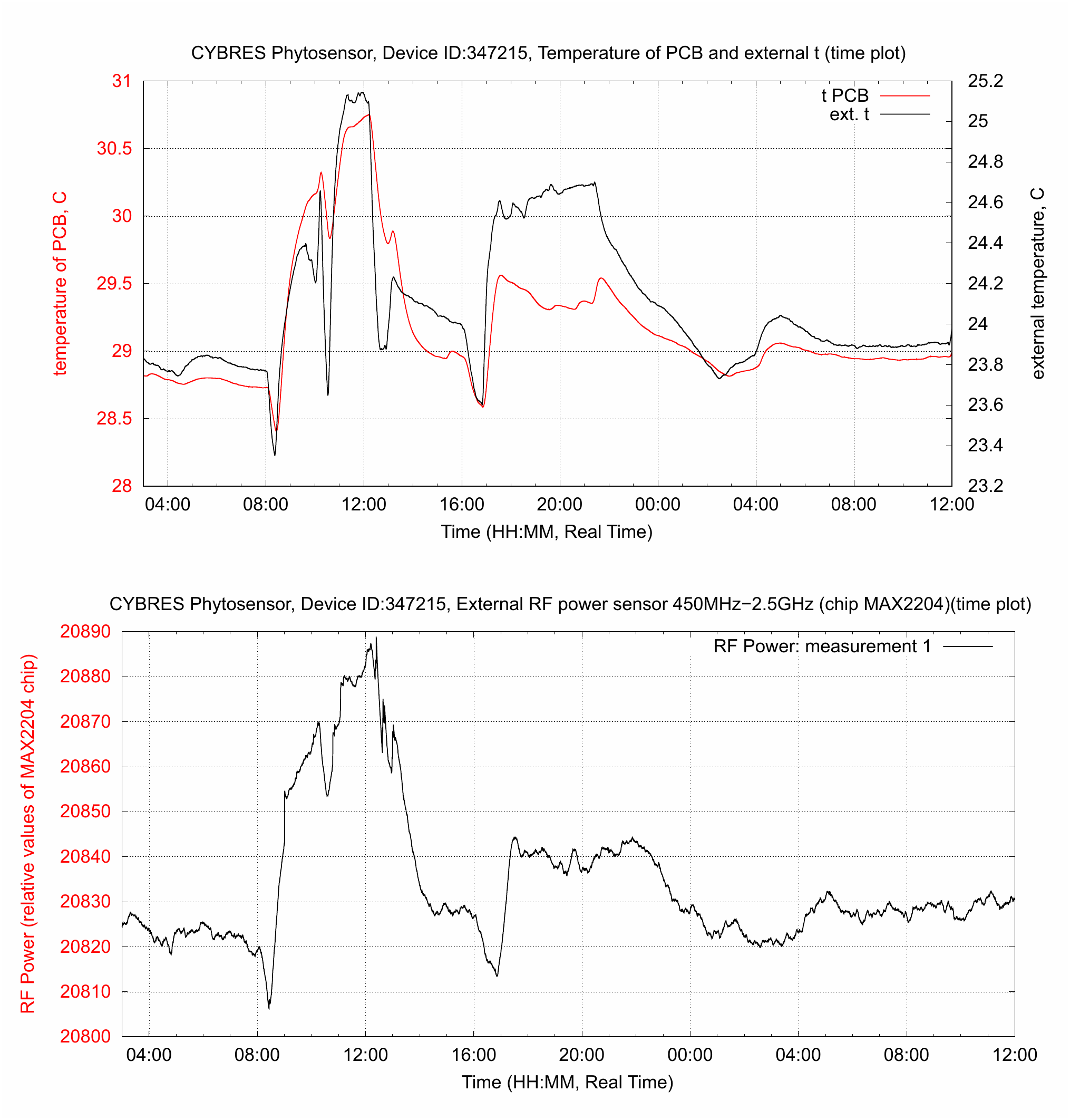}
    \caption{Test measurements: environmental data.\label{fig:testMeasurement3}}
\end{figure}

\begin{figure}[htp]
\centering
\subfigure{\includegraphics[width=.49\textwidth]{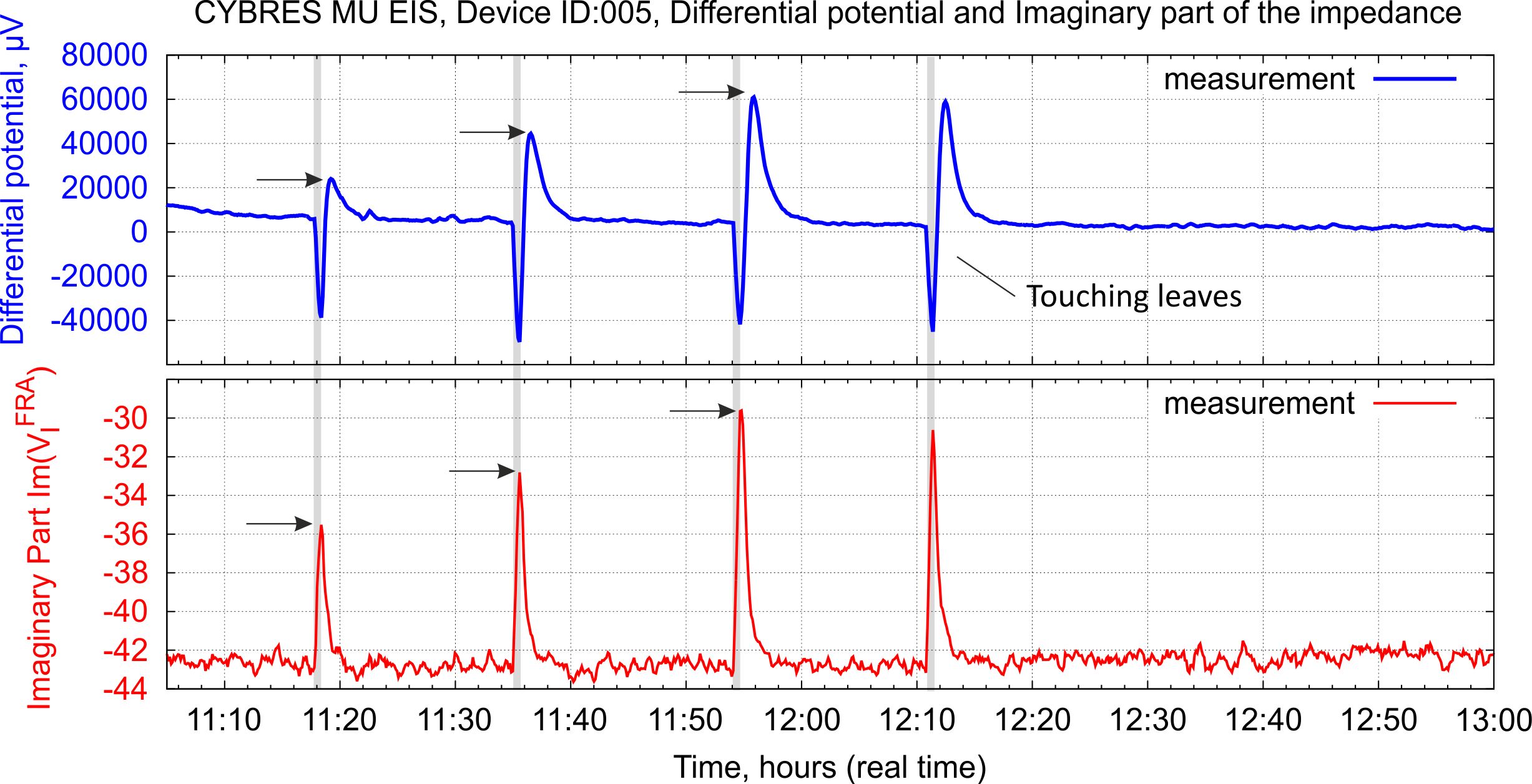}}
\caption{\small Example of electrophysiological responses of a plant on touch: (above) electric potential; (below) the imaginary part of tissue impedance at the frequency of 500 Hz. \label{fig:touch}}
\end{figure}

Figs. \ref{fig:testMeasurement1}-\ref{fig:testMeasurement3} demonstrate several plots of physiological and environmental data obtained from different devices. Fig. \ref{fig:touch} shows an example of interactions with users -- an electrophysiological reaction of plant on touch as electric action potentials and as tissue impedance. This electric response can be  used for phytoactuation purposes.

\section{Phytoactuation}

The phytosensor system is able to perform in-hardware and in-software real time signal processing by means of embedded numerical processors and real-time detectors. Examples of numerical processors are the mean, standard deviation or z score calculations. Examples of detectors are the time interval detector, the peak detector, the cyclical change detector, the noise levels detector, the gradient change detector, time detector and others \cite{CYBRES_UserManual}. Each processor and detector Dx is implemented as an independent module, see Fig. \ref{fig:detectors}. Detectors take signals from short-term, middle-term or long-term data pipes and can be configured by user for processing any of the existing data channels (data from real sensors such as potential measurements, electrophysiology, electrochemical data or external sensors. All pipes prepare their data automatically:
\begin{itemize}
  \item \textbf{short-term data:} each data sample is prepared with timing defined in 'system' $\rightarrow$ 'period between measurements' (usually in sec. time step);
  \item \textbf{middle-term data:} data samples are selected with time steps in minutes. It is implemented as 'selecting one sample of short-term data' when the short-term buffer is full (cycle of short-term data data collection);
  \item \textbf{long-term data:} data samples are selected with time steps in hours. It is implemented as 'selecting one sample of middle-term data' when the middle-term buffer is full (cycle of middle-term data data collection);
\end{itemize}
The size of buffer for all data pipes is controlled from the configuration file. Thus, different data pipes provide a possibility to analyze different time scales in the sensor data. Each enabled detector writes the result of detection into the output vector, see Fig. \ref{fig:detectors}.
\begin{figure}[htp]
\centering
\subfigure{\includegraphics[width=.49\textwidth]{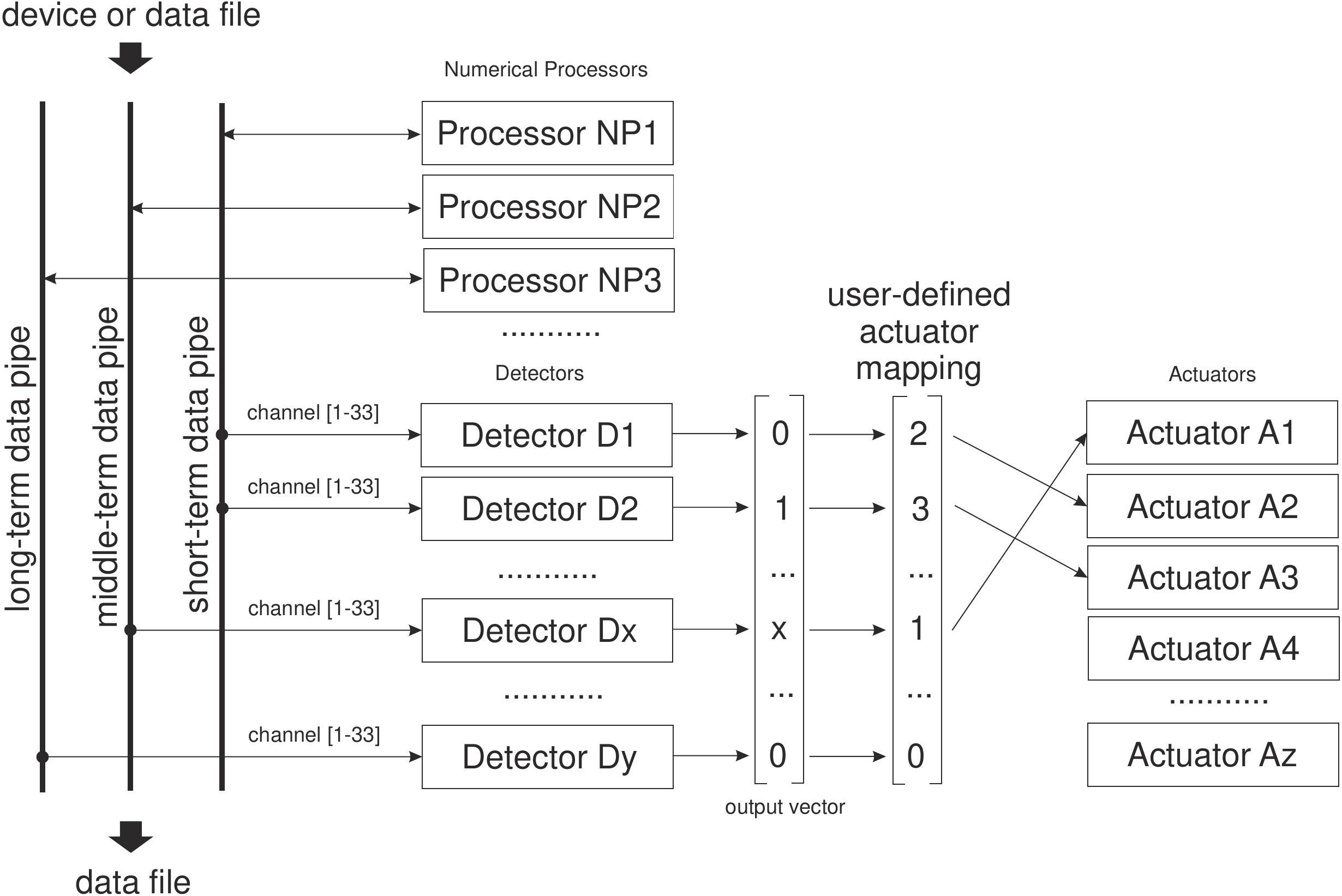}}
\caption{\small Schematic representation of the detector-actuator coupling. \label{fig:detectors}}
\end{figure}
\begin{figure}[htp]
\centering
\subfigure{\includegraphics[width=.49\textwidth]{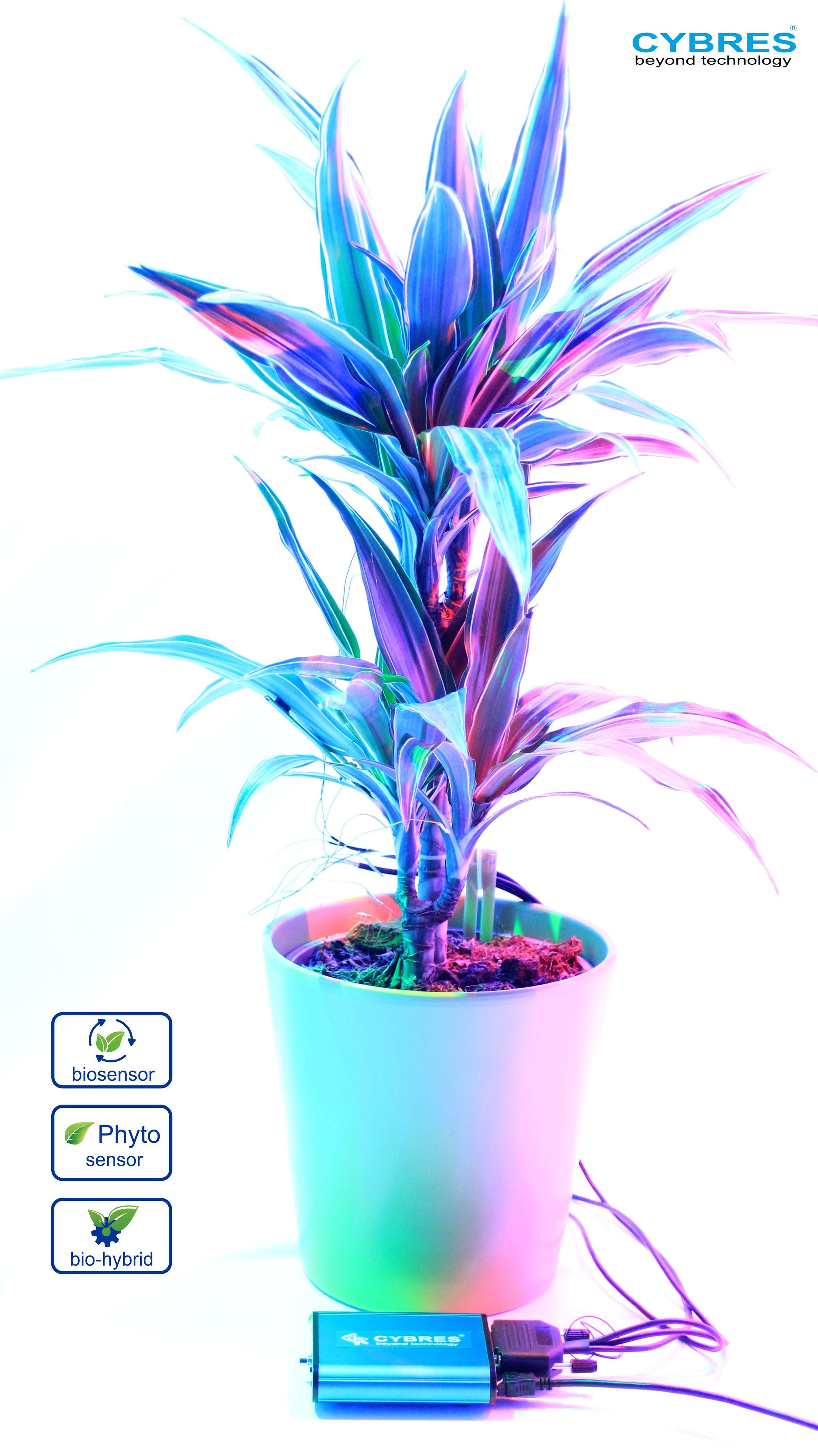}}
\caption{\small Hardware of the Phytosensor (MU3 unit + phytosensing electrodes) system. \label{fig:Phytosensor}}
\end{figure}
When the corresponding detector Dx detect the necessary changes in the signal (the 'true' condition), it writes '1' in the output vector, otherwise (the 'false' condition) it write '-1'. When the detector Dx was not executed at some step, it writes '0' in the output vector. Detectors can also write some numbers in the output vector, e.g. the noise level detector writes the noise level in the output vector. All detectors operate in parallel, they start when the corresponding data pipe issues the signal 'new data are ready'. The detector is switched off when its input data channel is configured as '0'. All detectors can be individually configured with one or several parameter. The list of available/optional actuators includes the following groups:
\begin{enumerate}
  \item \textbf{wavOut/mpegOut device:} play sound .wav/.mp3 file from position 'x' to position 'y';
  \item \textbf{sound device:} change volume, change right/left stereo balance;
  \item \textbf{MIDI device:} generate musical MIDI tones;
  \item \textbf{text-to-speech (TTS) device:} generate the voice message;
  \item \textbf{logical/probabilistic device:} to compute different logical/probabilistic operations and expressions from detectors;
  \item \textbf{adaptation mechanisms:} several instruments and methods to implement adaption in probabilistic networks;
  \item \textbf{RGB LED device:} turn on/off R,G,B components of LED connected to the MU system;
  \item \textbf{external physical devices connected to the MU system:} e.g. turn on/off lamps/pumps connected to the MU system (by sensing the ASCI commands in COM-port);
  \item \textbf{electrical stimulation device:} generate the electrical stimulation by the MU impedance measurement system;
  \item \textbf{external physical devices connected e.g. to USB port:} generic control of external devices;
  \item \textbf{'intelligent house' devices and systems:} on/off and parametric control of these devices;
	\item \textbf{send message to file:} write text message to file;
  \item \textbf{send message to IP port:} send text message to specific IP port in internet/intranet;
  \item \textbf{send message to twitter account:} send text message to specific twitter account.
\end{enumerate}
Each actuators is configured separately.

\section{Conclusion}

This work briefly introduced the phytosensor system, its main features, application cases and sensors. The result of these developments represents a compact system, see Fig.  \ref{fig:Phytosensor}, which can be used for different purposes and is commercially available. In this context, the phytosensor is an example of innovation 'from lab to the market', where scientific concepts are implemented in hardware and software and finally placed on global market.

\section{Acknowledgment}

This work is partially supported by the EU-H2020 Project 'FloraRobotica: Flora Robotica: Societies of Symbiotic Robot-Plant Bio-Hybrids as Social Architectural Artifacts', grant No: 640959, the FET Innovation Launchpad Project: 'BioHybrids: Biohybrid phytosensing system for plant-technology interactions in mixed-reality and smart-home systems', grant No: 945773, funded by European Commission.

\small
\IEEEtriggeratref{8}
%\bibliographystyle{unsrt}
%\bibliography{../bib-para,../bibl_sk,../own_bibl_sk}

\end{document}